\documentclass[%
 reprint,
superscriptaddress,
nofootinbib,
 amsmath,amssymb,
 aps,
floatfix,
twocolumn,
pre]{revtex4-2}

\usepackage{xcolor}
\usepackage{comment}
\usepackage{graphicx}
\usepackage{dcolumn}
\usepackage{bm}
\usepackage{upgreek}

\begin{document}

\preprint{APS/123-QED}

\title{Training Allostery-Inspired Mechanical Response in Disordered Elastic Networks
}

\author{Savannah D. Gowen}
 \email{gowen22s@uchicago.edu}
 \affiliation{
 Department of Physics and The James Franck and Enrico Fermi Institutes,
 University of  Chicago, Chicago,  IL  60637,  USA}

\date{\today}

\begin{abstract}
Disordered elastic networks are a model material system in which it is possible to achieve tunable and trainable functions. This work investigates the modification of local mechanical properties in disordered networks inspired by allosteric interactions in proteins: applying strain locally to a set of source nodes triggers a strain response at a distant set of target nodes. This is demonstrated first by using directed aging to modify the existing mechanical coupling between pairs of distant source and target nodes, and later as a means for inducing coupling between formerly isolated source-target pairs. The experimental results are compared with those predicted by simulations.
 
\end{abstract}

\maketitle


\section{Introduction}
The creation of novel artificial materials can often benefit from mimicking the robust abilities of materials found in living matter~\cite{bio_mat}. Rapid developments in materials science are increasingly pushing such capabilities through the development of biomimetic materials~\cite{biomimetics}, including materials that can heal~\cite{self_healing}, learn~\cite{learning}, and store memory~\cite{memory}. Included in this genre of materials are those that can be trained to perform unique functions that were not specifically designed into them at the outset~\cite{training}: One base material can be trained for a variety of different tasks. This allows for significant flexibility in the material function. Extensive research remains to determine the limits of adaptable function through training in non-living matter.

One biological example of unusual function comes from a phenomenon in protein dynamics called allostery:  the binding of a molecule at one site in a protein triggers the ability of a distant site to bind to another molecule~\cite{allostery}. This action-at-a-distance is one important way that proteins control their activity. Recent work has demonstrated the possibility of creating non-biological materials that mimic allosteric behavior~\cite{rocks,wyart_allostery,leibler_allostery}. Rocks \textit{et. al.} modeled elastic materials as disordered spring networks, which are composed of central-force spring bonds connected at nodes in a disordered network. Through the process of pruning selected bonds, they showed that distant sites on the network can be mechanically coupled to move either in-phase or out-of-phase with one another. Once designed, they showed that these tuned networks could be fabricated in the laboratory with physical materials. However, detailed computations of the network's elastic properties are required to design functionality into these networks; this becomes prohibitive in the design or fabrication of arbitrarily large networks. 

To avoid the use of direct computation, Pashine described an alternative method for achieving an allosteric-inspired function based on observations of a network's physical properties~\cite{nidhi}. In that work, bonds were pruned by observing stress-induced birefringence in a quasi-two-dimensional photo-elastic network. This technique was successfully implemented, however the procedure of bond pruning remains time-consuming in practice and is limited to materials that are optically responsive to force. 

 \begin{figure}[ht!]
\centering
\includegraphics[width=3.375in, height=3.3in]{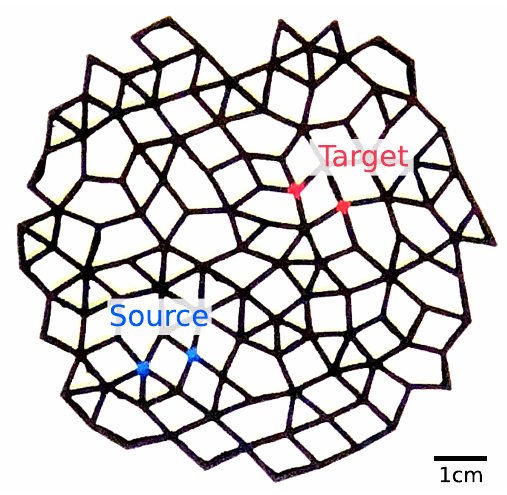}
\caption{
Disordered elastic foam network labeled with source and target node pairs.  
}
\label{elastic_network}
\end{figure}

Another approach, inspired by bond pruning, but which does not require the actual removal of any bonds or direct computation of the material's elastic properties, is referred to as directed aging~\cite{directed_aging}. This method is a form of training that takes advantage of a material’s innate ability to adapt to an applied stress load in time. This adaptation could be through the progressive weakening of bonds or the initiation of instabilities such as buckling to lower the energy of the directed state. Once aging is complete, the material should ideally respond according to how it was strained in the training process. 

The success of directed aging was demonstrated in the creation of materials with negative Poisson’s ratios.  Most naturally occurring materials have a positive Poisson’s ratio, $\nu > 0$, so that an applied strain in the material along one axis creates a strain of the opposite sign along its perpendicular axes. It is rare to find materials with $\nu < 0$, referred to as auxetic materials~\cite{auxetic}. It was demonstrated both in simulations and experiments that by aging a disordered elastic network under a compressive strain, one can selectively lower the network's bulk modulus with respect to its shear modulus~\cite{directed_aging,hexner}.  Because the Poisson’s ratio is a monotonic function of the ratio of the bulk to shear moduli, this can eventually lead to a material with a negative Poisson’s ratio. 

Although directed aging has been demonstrated as an effective means for modifying global material properties like the Poisson's ratio, it is not clear that directed aging can lead to localized function in physical materials in the laboratory. A bulk response requires aging contributions from bonds throughout the material, while in the case of mechanical allostery, the source and target responses are confined to local regions of the material. Although simulations have suggested that directed aging can be implemented to modify a network's local mechanical properties~\cite{hexner}, until now this has not been attempted in real materials.

In this work, I use directed aging as an unsupervised means of modifying the local mechanical coupling between separated sites in two-dimensional disordered networks. In the first example, I show that this type of training can be used to reduce the mechanical coupling between separated pairs of nodes that were initially coupled.  In a second example, I use training to induce a desired coupling between node pairs that were originally mechanically uncoupled. I compare these experimental outcomes to those previously observed in simulations.

\section{Experimental Protocol}

Disordered elastic networks are fabricated by laser cutting network designs derived from simulated two-dimensional jammed particle packings into a solid sheet of EVA foam. A set of two source and two target nodes are arbitrarily selected and labeled as shown in Fig.~\ref{elastic_network}. Initially, strain is applied to the source nodes as defined by $\epsilon_{\rm{s}}\equiv \Delta s / s$,
where $s$ is the initial distance between the two source nodes and $\Delta s \equiv s_{\rm{f}}-s $, the difference between final and initial node distances. The strain on the target nodes is measured in response as $\epsilon_{\rm{t}}\equiv \Delta t / t$ where t is the initial distance between the two target nodes and $\Delta t \equiv t_{\rm{f}}-t $, is the difference between the final and initial target node distances. The relative degree of coupling between distant node sites can be determined by measuring the 
ratio of the target to source strain: 
\begin{equation}
 \eta \equiv \epsilon_{\rm{t}} / \epsilon_{\rm{s}}
 \label{strain_ratio}
\end{equation}
Here $\eta \approx 0$ indicates uncoupled nodes, while $|\eta| > 0$ indicates coupled node pairs with a mechanical response that is either in-phase ($\eta > 0$) or out-of-phase ($\eta< 0$). 

Networks are trained via directed aging by applying a strain to the source and target node pairs for a fixed amount of time either statically or through (in-phase) cyclic driving (see Methods section for details). After a fixed aging time, $\tau_{\rm{age}}$, has elapsed, the constrained nodes are released, and the target strain is measured as a function of the varied source strain.

\begin{figure*}
\centering
\includegraphics[width=6.75in,height=2in]{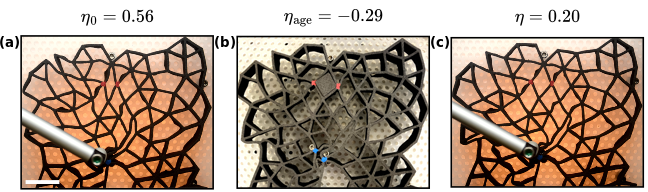}
\caption{Training protocol for suppressed node coupling via directed aging. \textbf{(a)} The initial strain-ratio, $\eta_{0}$, of a coupled source-target pair is measured. The source, identified by one node painted blue and another below the screw (painted green), is strained by approximately 0.75. The target, identified on the network by two red dots, contracts in response. The scale bar represents 2~cm. \textbf{(b)} To age the network, the source nodes are held compressed while the target is prevented from contracting (and is mildly stretched) by the insertion of a foam barrier resulting in a negative aging strain ratio, $\eta_{\rm{age}}$. The network remains statically in this configuration for a fixed time, $\tau_{\rm{age}}$. \textbf{(c)} The aged network is measured here shown at $\tau_{\rm{age}}=$ 24~hrs. Aging has led to visibly less contraction of the target when the same strain is applied to the source. 
}
\label{decoupling_ims}
\end{figure*}

\section{Results}

Experimentally modifying the local mechanics of these networks is challenging because there is no simple means of quantifying how stress is distributed in the material when strain is locally applied. Changes in geometry through node displacements, bond bending, and buckling are indicative of concentrated stress but are difficult to quantify.

The first experiments reported here take pairs of nodes that are already mechanically coupled and use directed aging to decouple their interactions. Networks are derived from jammed particle packings and there is a chance that some pairs of source and target nodes are coupled once created. I take advantage of this feature by selecting node pairs with existing interactions for training.  The second set of experiments uses aging techniques to establish mechanical coupling in pairs of nodes that were initially non-interacting. 

\subsection{Suppressing node coupling via directed aging}

Starting with source and target pairs that are mechanically coupled before any attempt to change the properties of the network, I apply a training protocol to decouple their motions. Once the coupled source and target nodes are determined, initial measurements of the strain ratio, $\eta_{0}$, are made by applying strains of approximately $0.25$, $0.50$, and $0.75$ $\pm 0.10$ to the source nodes and measuring the target strain response. One such measurement is shown in Fig.~\ref{decoupling_ims}a. 

To suppress the source-target interactions, I apply a large strain at the source while preventing the target nodes from reacting.  This is accomplished by placing a physical barrier at the target to arrest strain at those nodes, as seen in Fig.~\ref{decoupling_ims}b.  The strain at the source can be applied either statically (that is, by constraining the source nodes to a constant applied strain) or cyclically (by oscillating the strain at the source to a fixed maximum amplitude). These constraints are applied for a fixed aging time, $\tau_{\rm{age}}$, then released, and the response $\eta$ is measured. An example of this is shown in Fig.~\ref{decoupling_ims}c.  

\begin{figure}[h]
\centering
\includegraphics[width=3.375in,height=2.75in]{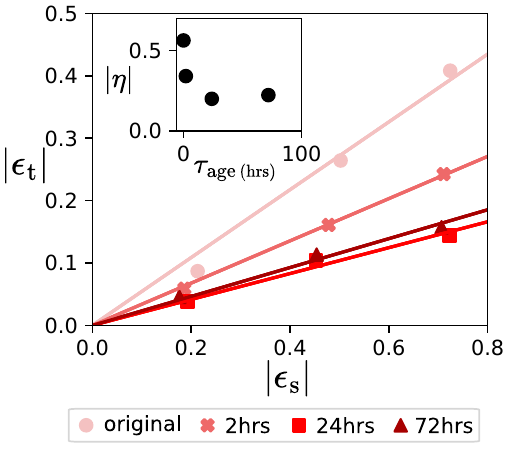}
\caption{Suppression of source-target strain coupling. The target strain, $\epsilon_{\rm{t}}$, is measured as a function of the source strain, $\epsilon_{\rm{s}}$, for varied aging times, $\tau_{\rm{age}}$, (represented by different colors and markers) for the network imaged in Fig.~\ref{decoupling_ims}. The strain ratio, $|\eta|$ is computed for the largest source strains and is shown in the inset as a function of $\tau_{\rm{age}}$. The strain ratio, $|\eta|$, decreases as a function of time with a signal suppression that saturates at or before 24 hours of static aging.
}
\label{decoupling_strains}
\end{figure}

Figure~\ref{decoupling_strains} shows the evolution of coupled source and target strains as a function of $\tau_{\rm{age}}$, for a single network. Source and target strains appear linearly coupled. The strain ratio, $\eta$, given by the slope of the line, decreases as a function of $\tau_{\rm{age}}$ until saturation at or before 24 hours. The inset shows $|\eta|$, the strain ratio, as a function of $\tau_{\rm{age}}$ at the largest source strain. These experiments were repeated for five distinct network geometries which featured different connectivity and void fraction, and for multiple coupled source-target pairs. After aging, 75\% of the networks showed a reduction in strain ratio by a factor of at least $0.5$, and all networks decreased by a factor of at least $0.25$. Less successful source-target pairs often had source and/or target nodes located on the boundary.

\begin{figure}
\centering
\includegraphics[width=3.375in,height=2.75in]{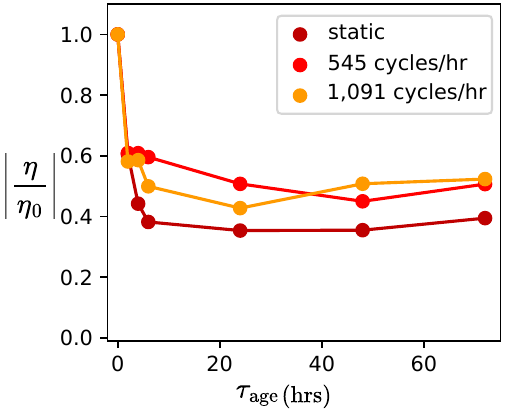}
\caption{Static versus cyclic training. The aged strain ratio, $\eta$, measured at the largest source strain is scaled by the initial strain ratio, $\eta_{0}$, for source and target nodes aged using three different protocols. One copy is aged statically while the other two alternate between static and cyclic activation as described in the text. All networks are aged for the same net aging time, $\tau_{\rm{age}}$. The number of training cycles appears to have a negligible effect on the training outcomes, however, networks held static for a longer duration show marginally better results.
}
\label{cyclic_static}
\end{figure}

The specific aging protocol has a relatively small effect on the training outcome. Figure~\ref{cyclic_static} compares the same network aged either statically or via cyclic driving. The aged strain ratio, $\eta$, scaled by the initial strain ratio, $\eta_{0}$, is plotted as a function of aging time, $\tau_{\rm{age}}$, for a network aged statically for the entire duration, one aged cyclically for 1/3 of that time and then held statically for the remainder, and another aged cyclically for 2/3 of that time and statically for the rest. Despite the number of cycles differing by a factor of two or more, there is very little difference in the aging behavior of the three networks. Rather, networks aged statically for a longer fraction of the time appear to yield slightly better results.

Suppressing local mechanical coupling through aging is demonstrated here for both in-phase and out-of-phase interactions between the source and target. Although most experiments include a target that is prevented from compressing when source strain is applied, it is also possible to prevent target nodes from expanding by applying constraints. Directed aging thus appears to provide a robust method for modifying local stress distributions in the material.

\subsection{Induced node coupling via directed aging}

\begin{figure*}
\centering
\includegraphics[width=6.75in,height=2in]{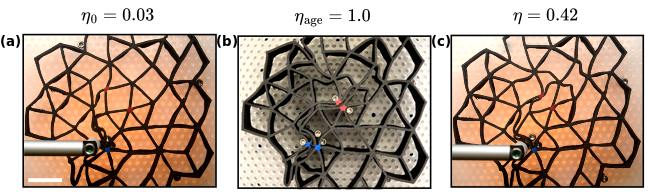}
\caption{Training protocol for inducing node coupling via directed aging. \textbf{(a)} The initial strain-ratio measurement of an uncoupled source-target pair is taken. When the source is strained by approximately 0.75, the distance between target nodes(red dots) is relatively unchanged. The strain ratio, $\eta_{0}$ is approximately 0. The scale bar represents 2~cm. \textbf{(b)} Directed aging is shown by applying compression statically to the source and target nodes for a set duration, $\tau_{\rm{age}}$. \textbf{(c)} The aged network is measured at $\tau_{\rm{age}}=$ 7~days. The aged target now contracts in response to an applied source strain, yielding a positive strain ratio.
}
\label{coupling_ims}
\end{figure*}

 Inducing coupling between distant nodes introduces a more formidable challenge than modifying existing ones. When strain is applied locally to a source chosen at random, surrounding nodes will be displaced leading to bond bending and buckling in the region surrounding the source. Previous work shows that there is a length scale, $\xi$, associated with the decay of local stress that depends on the network coordination~\cite{decay_length}. For directed aging to work effectively, stress applied to the source and target must affect a set of common bonds~\cite{nidhi,hexner}, and this is not possible if the distance between the source and target is too large, and/or if $\xi$ is too small. Because $\xi$ cannot easily be measured in the experiment, targets are chosen by straining the source and locating target nodes that are close by, but which remain unaffected when the source is strained.  Figure~\ref{coupling_ims}a shows one such example. The network is trained by applying a large negative (compressive) strain to both the source and target and allowing the network to age either statically [Fig~\ref{coupling_ims}b] or by cyclic driving. The rest of the network remains unperturbed for the duration of the training. The network is then remeasured after aging [Fig~\ref{coupling_ims}c]. 

\begin{figure}[h]
\centering
\includegraphics[width=3.375in,height=4.2in]{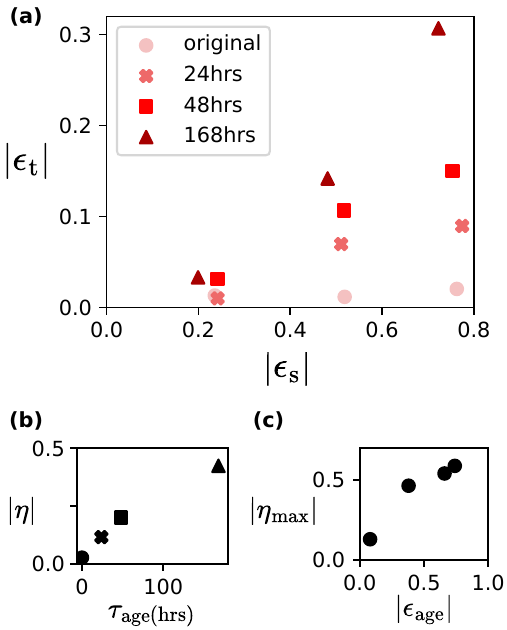}
\caption{Induced node coupling as a function of aging time, $\tau_{\rm{age}}$. \textbf{(a)} The target strain, $\epsilon_{\rm{t}}$, increases as a function of source strain, $\epsilon_{\rm{s}}$, with progressive aging time(indicated by different colors and markers) for the network imaged in Fig.~\ref{coupling_ims}. \textbf{(b)} The strain ratio, $|\eta|$, is measured at the largest source strain values and increases from a value of 0 (uncoupled nodes) to roughly 0.5 (coupled nodes) with $\tau_{\rm{age}}$. \textbf{(c)} The maximum strain ratio, $|\eta_{\rm{max}}|$, is measured as a function of the target aging strain, $\epsilon_{\rm{age}}$, for four copies of the network shown in Fig.~\ref{coupling_ims} aged at different strains. The largest induced strain ratio corresponds with the largest aging strain. 
}
\label{coupling_strains}
\end{figure}

Figure~\ref{coupling_strains}a shows the evolution of source and target strain as a function of training time, $\tau_{\rm{age}}$, for one example. Initially, the target strain, $\epsilon_{\rm{t}}$, remains constant as the source is strained such that the original data can be functionally described by $\epsilon_{\rm{t}}=0$, as one would expect for uncoupled nodes. As the aging time, $\tau_{\rm{age}}$, increases, we see that the source and target become increasingly coupled with $\epsilon_{\rm{t}}$ growing as a function of $\epsilon_{\rm{s}}$. This is further demonstrated by observing the strain ratio, $\eta$ as a function of  $\tau_{\rm{age}}$ at the largest source strains in Fig.~\ref{coupling_strains}b. Figure~\ref{coupling_strains}c shows the maximum output strain ratio, $|\eta_{\rm{max}}|$ for four copies of the same network aged at different imposed target strains. All four networks demonstrate successful coupling, however, the network with the largest output strain ratio was aged under the largest strain. This further motivates training at maximal strain values.

This protocol is repeated for six different network geometries and a variety of source-target pairs. The strain ratio for each uncoupled network is initially $0 \pm 0.05$. Successful coupling is defined when the strain ratio increases to $|\eta| \geq$ 0.1 for any of the applied source strains. Successful coupling was achieved for 73\% of experiments. It should be noted that unlike in the previous set of experiments, here strain interactions in the network can behave non-linearly such that it is possible to measure coupling at one value of the source strain but not necessarily at another. I generally observe that the largest coupling is present at the greatest applied source strain value, but not exclusively so.  

Some source-target pairs fail to couple the input and output response. Reasons for failure may include attempting to couple nodes that are sufficiently far from each other as compared to the decay length of local stresses, $\xi$. Failure is also more likely when the target is located too close to a constrained source node. In this case, some nearby bonds may be prevented from transmitting stress and thus impede node coupling. This may also come from the fact that stress around a point source in these systems can be highly anisotropic~\cite{decay_length}. Among other failure modes were those in which the target nodes were trained under large tensions to induce node expansion. These bonds appear to have undergone such significant plastic deformation that they became highly uncoupled from the rest of the network. Inducing target node expansion might still be accomplished with lower aging strains but further exploration is necessary. 

Networks trained statically or cyclically could both successfully couple source and target nodes. However, static aging consistently yielded a stronger coupling for the same aging time.

 \begin{figure}[h]
\centering
\includegraphics[width=3.375in,height=2.7in]{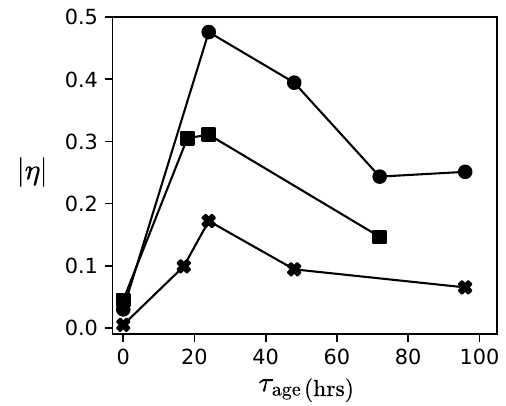}
\caption{Over-trained networks. The strain ratio, $|\eta|$, is shown as a function of aging time, $\tau_{\rm{age}}$, at high source strains for three different experiments (indicated by different markers). The curves shown with square and circular points are two copies of the same source-target pair, while the crosses represent another source-target pair. $\eta$ reaches a maximum value, or maximum coupling, at some intermediate time and then falls non-monotonically.
}
\label{over_training}
\end{figure}

\subsection{Over-training}
The training for allosteric response is not always monotonic in time. For over a quarter of experiments, if aged for too long, the coupling response can deteriorate. Thus, in some networks, $|\eta|$ was maximum at an intermediate time, as shown by a few examples in Fig.~\ref{over_training}. This response seems more common for networks containing bonds with high aspect ratios and is likely due to the viscoelastic properties of the foam. During aging, material struts under large stress loads buckle; sustaining this buckling for long periods can result in material fatigue which then drastically weakens the elastic response of the bond, rendering the training less effective. This effect is not unique to our system~\cite{chooki}. This decrease or plateau in the strain ratio is observed more frequently in networks aged for long durations, and should be observed in all networks if aged for long enough. This phenomenon is evidence that one can over-train a material, at which point, aging adopts the negative connotations that one may have naturally associated with the term.

\section{Discussion}

These experiments both validate the computational findings of Hexner \textit{et al.}~\cite{hexner}, and introduce a few notable differences. The simulations model elastic materials as disordered, central-force spring networks. The networks are aged by performing an energy minimization algorithm in which the rest length of each bond evolves with the stress applied to that bond when subjected to an applied strain. In their study, strain is applied cyclically while the networks age as a means of training the output signal for a range of input strains, rather than a single one. In these experiments, for both coupling and de-coupling nodes, I find that static training is equally or more effective than cyclic driving. The majority of networks aged statically at a single large value of $\eta_{\rm{age}}$ display the trained response at smaller strains as well. These results may imply that cyclic driving could be unnecessary to locally train networks over a range of source strain values. 

The coupled interactions achieved in these experiments were significantly smaller in magnitude than those from simulations. In simulations, a large percentage of source-target pairs could be trained to strain ratios of $\eta \approx 1$~\cite{hexner}. In experiment, the highest observed output strain ratio was $|\eta| = 0.64$. However, our results cannot be directly compared because the aging strains used in these experiments were larger than those considered in simulations, with a typical aging strain around $\epsilon_{\rm{s}} \approx 0.75$ versus 0.5 in simulations. The differences in strain ratio output are likely due to the more complex adaptive behaviors of physical networks that are not accounted for in the computational models. For example, bond bending, the change in the angle at which bonds meet at a node, has been observed to be one critical mechanism for geometry changes in experimentally aged networks. Bond buckling also represents an instability which once plastically set in the material, has the effect of severely decreasing the elastic energy stored in a bond. These same mechanisms most likely account for our observations of non-monotonic aging output, or over-training, which are not observed in the simulation results reported. 

Additionally, Hexner \textit{et al.} suggests that the inclusion of “repeaters”, randomly chosen nodes throughout the network that are strained in addition to the source and target, could ameliorate the failure rate in training long-range targets~\cite{hexner}. In some cases, duplicating failed experiments with the inclusion of repeaters does lead to an induced coupling, but not consistently so.  The dependence of coupling on the source to target distance, and the efficacy of repeaters in increasing the range of source-target interactions remain to be further investigated in these experimental systems.

Another feature of real materials that is not present in idealized spring models is the inclusion of pre-stress in bonds. Although the initial network is in static equilibrium, stresses may still exist in force balance with one another. It has been shown that the inclusion of even a small amount of pre-stress in spring models can have significant implications on the training outcomes~\cite{ayanna,liu_prestress}. This could also contribute to disparities in experimental and simulation results.

This work demonstrates experimentally that local material properties can be trained via directed aging. Mechanical coupling was reduced by at least 1/4 of the initial strain ratio for source-target pairs in 100\% of the de-coupling experiments. In the second set of experiments, although source-target nodes could not always be coupled to $|\eta| \geq 0.1$, the success rate was comparable to values previously reported. In experiments in which the same type of allosteric mechanical interactions were induced with bond pruning, Pashine reports success rates between approximately 70\%-90\% for the same strain ratio values~\cite{nidhi}. The inability to couple nodes through pruning was attributed to bond-bending and non-linearity in the stress-strain response of the material. These non-linear effects are likely significant to these experimental results as well. However, in simulations of directed aging there is a 10\% failure rate even without these added complexities included in the computational model~\cite{hexner}. In this work, approximately 75\% of source-target pairs are coupled by atleast 10\% after aging. The ability to train a relatively large fraction of source-target node pairs without the added insight of computation in experiment evidences the robust ability of disordered materials to adapt, and the promise of directed aging as a material design technique.

\section{Conclusions}

The experiments reported here demonstrate that local mechanical coupling between distant pairs of nodes can be modified or induced via directed aging in disordered elastic network materials. The ability to tune a material’s mechanical properties locally without large-scale computations or design presents a promising new direction for materials development that takes inspiration from modes of response in biological systems. 

In disordered foam networks, by forcing source and target bonds into geometric configurations in which the source and target are actively strained and allowing them to age, it is believed that the system evolves to a state in which these configurations become energetically favorable through the weakening of bonds under stress.  This presumed material adaptation is dependent on the viscoelastic nature of the underlying foam material. More work remains to characterize the limitations of aging in these types of material networks. Cases of failed node coupling might inspire further exploration for more effective training protocols. For example, varying ambient temperature conditions could allow a more rapid aging response and longer memory of the training.

Finally, it would also be interesting to analyze how these results change when the material of the network is changed. Materials that are, for example, strain hardening, may introduce other novel training outcomes.

\section{Methods}
In these experiments, we use disordered network structures derived from 2D computer-simulated jammed particle packings. Each node corresponds to the center of a circular particle while bonds indicate particle contacts where circles overlap. This structure is then laser cut using a Universal Laser Systems Ultra X6000 into a solid sheet of EVA foam that is 0.5 inches thick. Source and target node pairs are labeled with red and blue paint to facilitate strain measurements. Network boundaries are constrained on a perforated platform with obstacles to inhibit global translation or rotation of the material while a strain is being applied. 
To apply source and target strain, one source node is fixed while the strain is applied to the other using a linear actuator with feedback (Actuonix Motion Devices Inc.). Source and target responses are captured for various source strains using a Nikon D7000 DSLR camera. Training via directed aging is applied cyclically by synchronizing actuators to apply strain at both the source and target nodes repeatedly, or statically by constraining the source and target nodes as desired for a fixed amount of time. Cyclic actuation is driven at a frequency of 450~mHz using a motor driver powered by an Arduino. For all measurements, the source nodes are forced back to their approximate initial positions. 

\section{Acknowledgments}
I would like to thank Nidhi Pashine for helpful advice and insightful discussions throughout the development of these experiments. I would also like to thank Samar Alqatari and Varda Hagh for their perspectives and guidance on the simulations. Finally, I would like to especially acknowledge the contributions of Sidney R. Nagel as an advisor and mentor throughout the development of this work. This research was supported by the University of Chicago Materials Research Science and Engineering Center, NSF-MRSEC program under award NSF-DMR 2011854.

\bibliography{main}

\end{document}